\documentclass[aps,prl,twocolumn,showpacs]{revtex4-1}
\usepackage{graphicx}

\usepackage{color}

\begin{document}


\title{Rotational Symmetry Breaking in a Trigonal superconductor Nb-doped Bi$_{2}$Se$_{3}$}
\author{Tomoya Asaba$^1$, B.J. Lawson$^1$, Colin Tinsman$^1$, Lu Chen$^1$, Paul Corbae$^1$, Gang Li$^1$, Y. Qiu$^2$, Y.S. Hor$^2$, Liang Fu$^3$, and Lu Li$^1$}

\affiliation{
$^1$Department of Physics, University of Michigan, Ann Arbor, MI  48109 USA \\
$^2$Department of Physics, Missouri University of Science and Technology, Rolla, MO 65409 USA\\
$^3$Department of Physics, Massachusetts Institute of Technology, Cambridge, MA 02309 USA
}
\date{\today}
\begin{abstract}
The search for unconventional superconductivity has been focused on materials with strong spin-orbit coupling and unique crystal lattices. Doped bismuth selenide (Bi$_2$Se$_3$) is a strong candidate given the topological insulator nature of the parent compound and its triangular lattice. The coupling between the physical properties in the superconducting state and its underlying crystal symmetry is a crucial test for unconventional superconductivity. In this paper, we report direct evidence that the superconducting magnetic response couples strongly to the underlying 3-fold crystal symmetry in the recently discovered superconductor with trigonal crystal structure, niobium (Nb)-doped bismuth selenide (Bi$_2$Se$_3$). More importantly, we observed that the magnetic response is greatly enhanced along one preferred direction spontaneously breaking the rotational symmetry. Instead of a simple 3-fold crystalline symmetry, the superconducting hysteresis loop shows dominating 2-fold and 4-fold symmetry. This observation confirms the breaking of the rotational symmetry and indicates the presence of nematic order in the superconducting ground state of Nb-doped Bi$_2$Se$_3$. Further, heat capacity measurements display an exponential decay in superconducting state and suggest that there is no line node in the superconducting gap. These observations provide strong evidence of odd-parity topological superconductivity.
\end{abstract}

\maketitle

Unconventional superconductors are characterized by superconducting order parameters that are non-invariant under crystal symmetry operations. When the order parameter is single-component, this non-invariance is manifested solely in the phase of the superconducting wavefunction, and can only be detected by phase-sensitive measurements. On the other hand, when the order parameter is multi-component, the magnitude of the superconducting gap can be different along symmetry-related crystallographic directions. The gap anisotropy directly leads to thermodynamic property of the superconducting state that spontaneously breaks the crystal rotational symmetry of the normal state. However, direct thermodynamic signature of rotational symmetry breaking due to superconductivity has not been found in any crystals.   
 
Bismuth selenide (Bi$_2$Se$_3$) makes the best material system to explore for unconventional superconductivity. The strong spin-orbit coupling in the triangular lattice has led to the topological insulating ground state in Bi$_2$Se$_3$~\cite{RMPHasan,RMPZhang}.  Doping with metallic element such as copper (Cu) and strontium (Sr) made it superconducting ~\cite{Hor2010, Wray2010, FuBerg,PRLHeatCapacityAndo, Liu2015, LawsonPRL, LawsonPRB,Sasaki, Lahoud}. We report here the first direct observation of rotational symmetry breaking in the superconducting property of Nb-doped Bi$_2$Se$_3$~\cite{QiuHor}, a new member of superconducting doped topological insulators in addition to Cu- and Sr-doped Bi$_2$Se$_3$.  Possible odd-parity pairing symmetries in doped Bi$_2$Se$_3$, favored by strong spin-orbit interactions, have been theoretically proposed and classified according to the representations of the $D_{3d}$ point group~\cite{FuBerg}. 
Among them, only  the odd-parity pairing in the two-dimensional $E_u$ representation gives rise to a nematic superconductor with broken rotational symmetry~\cite{FuPRB2014}.  A recent nuclear magnetic resonance experiment on Cu$_x$Bi$_2$Se$_3$ reveals a twofold in-plane anisotropy in the spin susceptibility of the superconducting state~\cite{Zeng}, providing spectroscopic evidence for the $E_u$ pairing. The pairing symmetries of Nb- and Sr-doped Bi$_2$Se$_3$ remain unknown.  

We applied torque magnetometry to map the complete angular dependence of the in-plane magnetic anisotropy in Nb-doped Bi$_2$Se$_3$. The in-plane magnetization displays the field dependent hysteresis characteristic of a type-II superconductor. The observed hysteresis shows a large twofold anisotropy, which reveals the broken rotational symmetry in Nb$_x$Bi$_2$Se$_3$. Our work establishes torque magnetometry as a new and powerful method for discovering nematic  superconductivity~\cite{MatsudaNature, MatsudaScience}.

\begin{figure}[h]
\includegraphics[width=3.3 in]{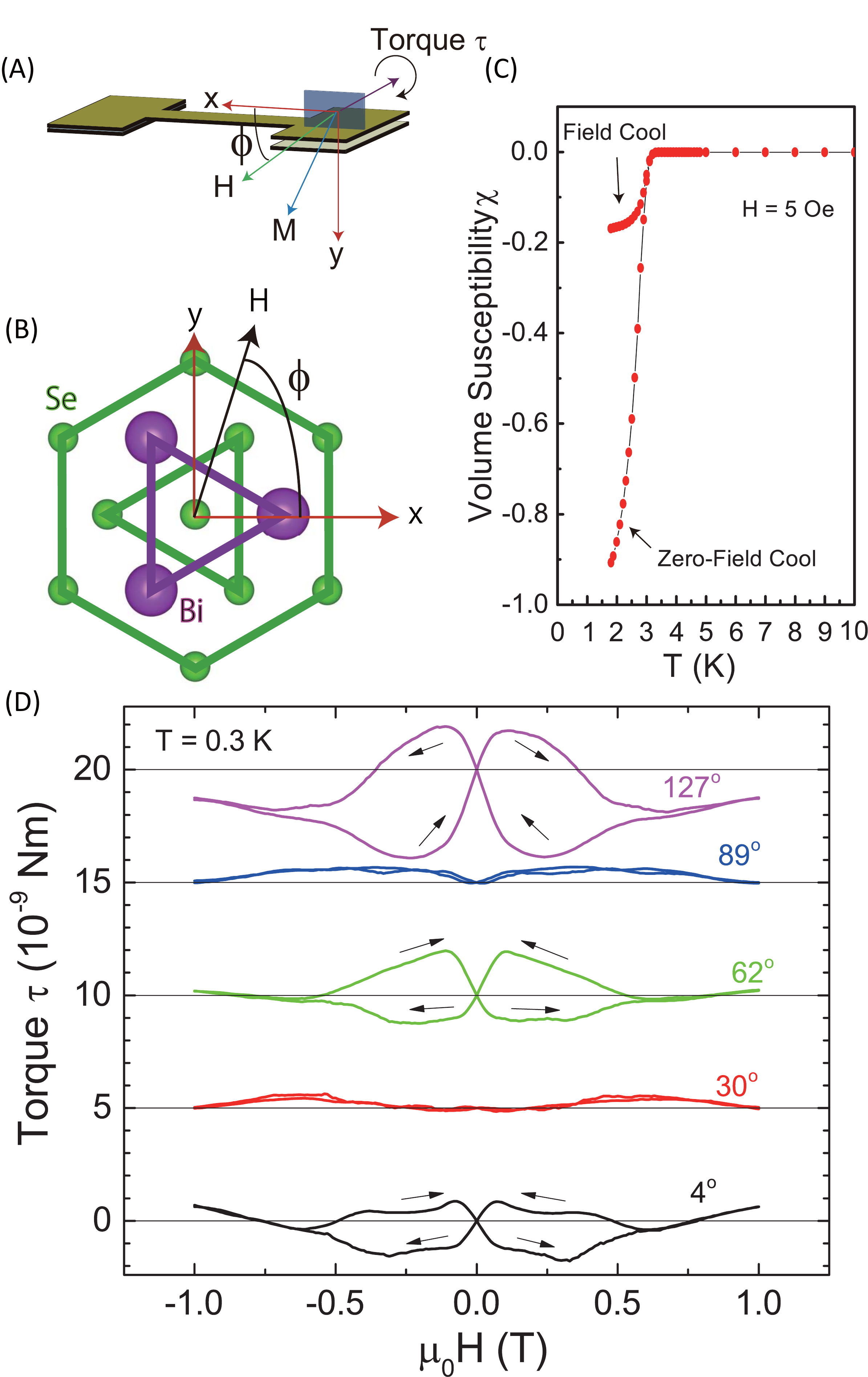}
\caption{\label{figtorque} 
{\bf Experimental setup, sample orientation  and torque curves of Nb$_{x}$Bi$_{2}$Se$_{3}$.} (color online)
(A) Schematic sketch of torque magnetometry under in-plane field rotation. The magnetic field is applied in-plane. The azimuthal angle $\phi$ is defined as the angle between the external magnetic field and the cantilever arm ($x$-axis). Torque $\vec{\tau}$=$\mu_{0}V\vec{M}\times\vec{H}$  is tracked by measuring the capacitance between the cantilever and the gold film beneath it. (B) Crystal structure of Nb$_{x}$Bi$_{2}$Se$_{3}$ viewed down the crystalline $\hat{c}$-axis. (C) Meissner effect from the sample. The volume magnetic susceptibility reaches close to -1, indicating a fully superconducting volume. (D) Selected torque curves at 0.3 K with external magnetic field between -1 and 1 T. The magnitude of the hysteresis loop is maximum at around $120^{\circ}$ and is nearly zero at $30^{\circ}$ and $90^{\circ}$.
}
\end{figure}

We used torque magnetometry to measure the superconducting hysteresis loop and magnetization of Nb-doped Bi$_{2}$Se$_{3}$.
Magnetic torque is given by $\vec{\tau}$=$\mu_{0}$V$\vec{M} \times \vec{H}$. Here $V$ is the volume of the sample, $\vec{H}$ is the external magnetic field, and $\vec{M}$ is the magnetization of the sample, given by the derivative of the free energy with respect to the external field $H$. Torque magnetometry is thus a thermodynamic probe that measures the free energy in a sample. 
The torque is measured by mounting the sample standing on its edge in order to keep the external field in the ab-plane (see Fig. \ref{figtorque}(A)).  We then rotated the cantilever. This measures the in-plane anisotropy of the sample's magnetic properties in the superconducting state. Fig. \ref{figtorque}(B) shows the crystal structure of Nb-doped Bi$_{2}$Se$_{3}$ looking down the hexagonal axis. As shown in the figure, the external field is in the hexagonal plane. The azimuthal angle, $\phi$, is the angle between the external magnetic field and the $x$-axis defined along the cantilever arm. Based on the X-ray diffraction pattern of this particular sample, we find that $\phi = 0^\circ$, $60^\circ$, and $120^\circ$ corresponds to the in-plane mirror axis of the crystal, as shown in Fig. \ref{figtorque}(B).

The samples of Nb-doped Bi$_{2}$Se$_{3}$ used in our experiment have superconducting volume close to 100\%, as shown by the volume magnetic susceptibility which approaches -1 in the zero-field cooled run (see Fig. \ref{figtorque}(C)). This is much higher than that of Cu-doped Bi$_2$Se$_3$ \cite{LawsonPRL,PRLHeatCapacityAndo}. 

The measured torque shows a strong superconducting signal. Figure \ref{figtorque}(D) shows some examples of the magnetic torque from the sample. The torque $\tau$ is plotted as a function of external magnetic field at temperate $T$ = 0.3 K. We swept the field up and down from -1 T to 1 T to measure the entire superconducting hysteresis loop. Arrows along the curve show the field sweep direction. The $\tau-H$ loop is a signature of the strong flux pinning characteristic of type-II superconductors. The pinned flux lines form a vortex solid, and the flux density inside the superconductor always resists the change of the applied magnetic field. A simple analysis based on the Bean model shows that the hysteresis of the magnetization gives a direct measurement of the superconducting critical current density in the mixed state of type-II superconductors (see supplemental materials). 

We note that the superconducting hysteresis loops at the selected angles have a clear angular dependence in $\phi$. The magnitude of the hysteresis loop reaches a maximum at around 60$^{\circ}$ and is nearly zero at $30^{\circ}$ and $90^{\circ}$. The variation of the hysteresis loop size is the first indication of anomalous in-plane symmetry in Nb-doped Bi$_2$Se$_3$.

\begin{figure}[h]
\includegraphics[width=3.3 in]{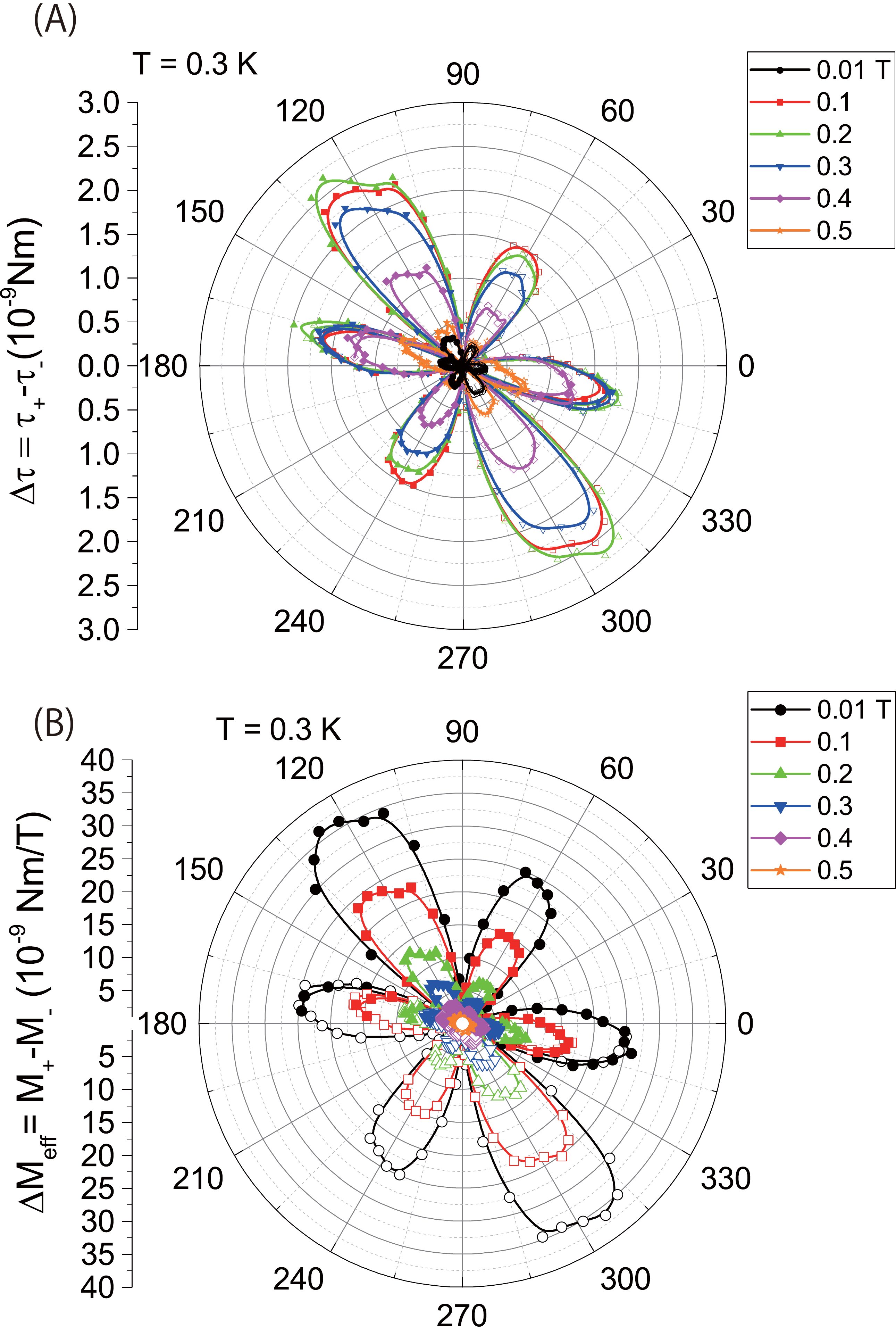}
\caption{\label{figpolar} 
{\bf Polar plot of hysteresis loop magnitude and spontaneous magnetization.} (color online)
(A) Polar plot of hysteresis loop magnitude $\Delta \tau = \tau_{+}-\tau_{-}$. $\tau_+$ is the torque signal from the up-sweep of the magnetic field, and $\tau_-$ is the torque signal from the down-sweep of the magnetic field. Angle $\phi$ is defined as the tilt angle between the positive $x$-axis and the magnetic field direction. Reversing the sign of the magnetic field is equivalent to rotating the cantilever 180$^{\circ}$. We thus used the negative field values to complete the 360$^{\circ}$ angular dependence (open circles). The plot of $\Delta \tau$ vs. tilt angle $\phi$ shows a periodic vanishing every 60$^{\circ}$. (B) Polar plot of spontaneous magnetization $\Delta M_{eff} = M_{+}-M_{-}$, where $M_{\pm}=\tau_{\pm}/\mu_0H$.
}
\end{figure}

To further illustrate the angular dependence of the superconducting hysteresis loop, we mapped the magnitude of the hysteresis loop as a function of $\phi$. To find the magnitude of the hysteresis loop, we took the difference of the sample's magnetic torque from the $H$-increasing sweep ($\tau_+$) and $H$-decreasing sweep ($\tau_-$). We denote this magnitude as $\Delta\tau=\tau_{+}-\tau_{-}$. Figure \ref{figpolar}(A) shows the absolute value of the hysteresis magnitude in a polar plot where $\Delta\tau$ is plotted against the azimuthal angle, $\phi$. $\Delta\tau$ was taken at different values of $H$. The sample was measured at $T$ = 0.3 K over an angular range of 200$^{\circ}$. We note that reversing the sign of the magnetic field is equivalent to rotating the cantilever 180$^{\circ}$. We thus used the negative field sweep data to complete the 360$^{\circ}$ angular dependence. The six-fold symmetry is clearly demonstrated by the observation that $\Delta\tau$ goes to zero every 60$^{\circ}$. This confirms the $H$-symmetric nature of the magnetic torque from the sample. Figure \ref{figpolar}(B) is the polar plot of spontaneous magnetization, $\Delta M_{eff} = M_{+} - M_{-}$, where $M_{\pm} = \tau_{\pm}/\mu_{0}H$.

\begin{figure}[h]
\includegraphics[width=3.3 in]{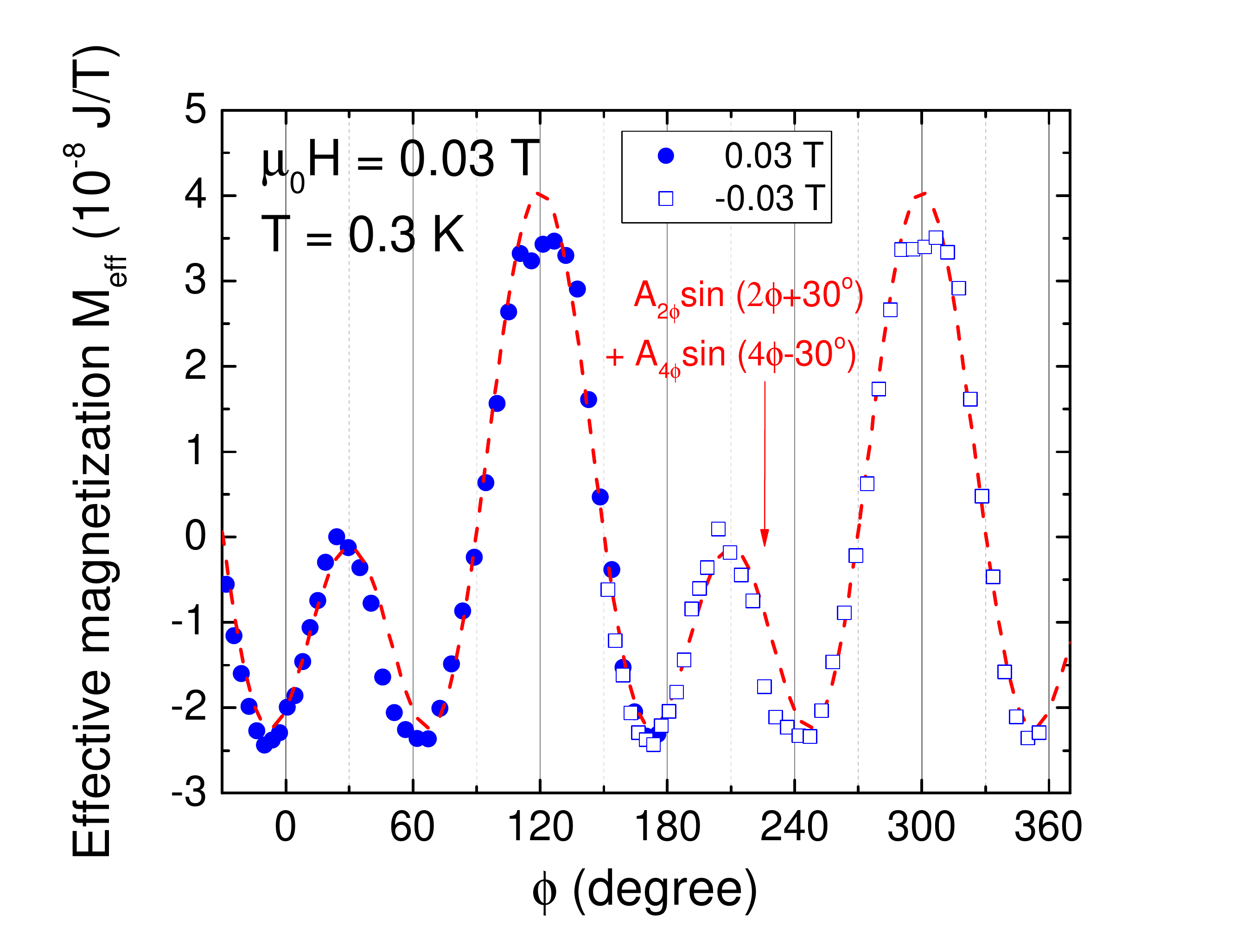}
\caption{\label{figfit} 
{\bf The angular dependence of spontaneous effective magnetization.} (color online)
The data were taken at 0.3 K at 0.03 T. Data taken from the positive field sweep are plotted as filled circles, and data from the negative field sweep are plotted as open circles. The plot of spontaneous effective magnetization, $\Delta M = M_{+}-M_{-}$, vs angle $\phi$ agrees with the fitting function $f(\phi) = A_{2\phi}\sin(2\phi+30^\circ)+A_{4\phi}sin(4\phi-30^\circ))$.  For this trace, $A_{2\phi} =-2.08\times 10^{-8} J/T$, and $A_{2\phi} = +1.96\times 10^{-8} J/T$.
} 
\end{figure}

The most important observation is the angular symmetry of the hysteresis loop near zero field. Figure \ref{figfit} displays the angular dependence of the effective magnetization hysteresis,  $\Delta M = {M_+-M_-} = \frac{\Delta \tau}{\mu_0H}$ at $\mu_{0}H$ = 0.03 T. The fitting function, $\Delta M\ =  A_{2\phi}\sin(2\phi+30^\circ)+A_{4\phi}sin(4\phi-30^\circ))$, shows the symmetry of the hysteresis loops.  The existence of the 2-fold symmetric $\sin{2\phi}$ contribution, as well as the 4-fold symmetric $\sin{4\phi}$ term, suggests the rotational symmetry breaking, i.e.,  nematic order, in superconducting Nb-doped Bi$_{2}$Se$_{3}$. 
   
\begin{figure}[h]
\includegraphics[width=3.3 in]{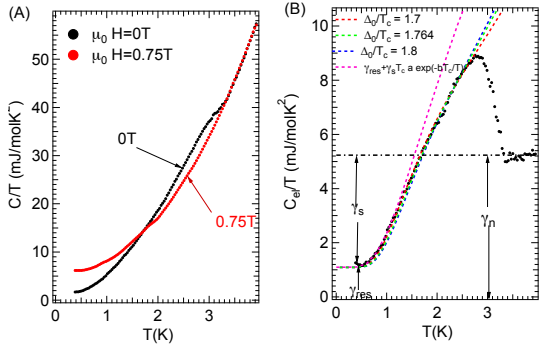}
\caption{\label{figCT} 
{\bf Fully gapped bulk superconductivity reveal by heat capacity measurement in Nb-doped Bi$_2$Se$_3$} (color online)
(A) Sample heat capacity $C$ is displayed as the ratio of $C$ and temperature $T$ plotted against $T$. The zero field curve is compared with the $\mu_0H = 0.75$ T curve. The 0.75 T curve provides us a method to define the phonon contribution $C_{ph}$. (B) The electronic part of heat capacity $C_{el}$ is shown as $\frac{C_{el}}{T}$ vs. $T$. A clear kink is observed at the superconducting transition temperature $T_c$. Near $T\sim0$, the curve approaches a finite value in $C_{el}/T$ and gives a measurement of the non-superconducting volume fraction around 20\% in this sample. As $T$ increases from the base temperature, $C_{el}/T$ follows the exponential curve(dashed pink line), as expected from a fully gapped superconductor. Numerical calculation of $C_{el}$ in BCS superconductors are shown with the only parameter $\alpha \equiv \frac{k_BT_c}{\Delta}$, where $\Delta$ is the superconducting gap. For the overall trace, $\alpha = 1.76$ trace gives the best fit of the heat capacity trace below $T_c$.
} 
\end{figure}

Finally, our heat capacity measurement on Nb-doped Bi$_2$Se$_3$ shows a fully gapped bulk superconductivity  (Fig. \ref{figCT}(A)). For the same sample, we measured the heat capacity $C$ at selected $T$ between 0.4 K and 20 K. Fig. \ref{figCT}(A) shows $\frac{C}{T}$ in the superconducting state at $\mu_0H$ = 0 T as well as at 0.75 T,  right above the closing of the hysteresis loop at base $T$. We note that above 4 K, the heat capacity $C$ at 0 T is the same as that at 0.75 T, within measurement errors. Therefore, we use the 0.75 T heat capacity curve as the normal state heat capacity $C_{n}$. From this we determined the phonon contribution following the same practice as the early work on Cu-doped Bi$_2$Se$_3$~\cite{PRLHeatCapacityAndo}:
\begin{equation}
C_n = C_{el}^n + C_{ph} = \gamma_nT + aT^3 + bT^5
\end{equation}
where the electronic heat capacity $C_{el}^n = \gamma_nT$ is for the normal state. Subtracting the phonon heat capacity, $C_{ph}$, we infer the superconducting state electronic heat capacity $C_{el}$ at 0 T, which is plotted as $\frac{C_{el}}{T}$ vs. $T$ in Fig. \ref{figCT}(B). The heat capacity at 0 T shows an exponential decay as $T$ drops to base temperature. This is a signature of a fully gapped, nodeless superconducting state. 

We note that $\frac{C_{el}}{T}$ trace approaches a finite value $\gamma_{res}$ near the base temperature. This suggests a partial non-superconducting volume in the crystal of about $\frac{\gamma_{res}}{\gamma_n}\sim$ $<$20\%. This value is consistent with the superconducting volume we determined from the Meissner effect. 

The exponential decay, rather than a typical power-law dependence, observed in $C_{el}-T$ is indicative of a nodeless superconducting gap. 
Other heat capacity papers on unconventional superconductors fairly recognize that a high order power-law can be hard to distinguish from an exponential decay at low temperatures ~\cite{Sasaki}. For line nodes, $\frac{C_{el}}{T}$ should fall linearly with $T$, and for point nodes, $\frac{C_{el}}{T}$ should go as $T^{2}$ ~\cite{Sigrist}. Line nodes, as have been seen in other unconventional superconductors ~\cite{MaoSRO, Hasselbach, Movshovich}, are clearly ruled out by our data. The absence of line nodes is consistent with odd-parity superconductivity in doped Bi$_2$Se$_3$~\cite{FuPRB2014, FuTSC2015}.

{\it Discussion} Our study of the magnetic torque demonstrates a symmetric 6-fold vanishing of the superconducting hysteresis. Magnetic torque is sensitive to the magnetic anisotropy of the superconducting signal, and the hysteretic $M-H$ loops arise from the flux pinning of the vortex solids in the superconducting state. Therefore, the superconducting diamagnetic signal, which comes from the vortex solid, prefers to align with the 3 mirror planes of the triangular lattice. The diamagnetic signal is maximum when the magnetic field aligns with these preferred directions. The magnetic torque $\vec{\tau} = \mu_0\vec{M} \times \vec{H}$ vanishes between these preferred directions when the $M$ vector is exactly collinear (parallel or antiparallel) of the applied field $H$.

The hysteresis loop is greatly enhanced along one direction. This suggests that the nematic order is a spontaneous symmetry breaking in the superconducting state. As pointed in ref. \cite{FuPRB2014}, the nematic order verifies the two-component nature of the superconducting order parameter. Thus an odd-parity superconducting order is very likely to exist in the ground state, which creates promise for topological superconductivity in Nb-doped Bi$_2$Se$_3$.

We note that early torque studies on high T$_c$ cuprates, untwinned YBa$_2$Cu$_3$O$_7$ \cite{IshidaYBCO} and Tl$_2$Ba$_2$CuO$_{6+\delta}$ \cite{WilleminTBCO}, also show in-plane vanishing of the superconducting hysteresis along their crystal axes. A similar measurement on NbN and NbSe$_2$, which are classical superconductors with s-wave symmetry \cite{WilleminTBCO}, showed no in-plane vanishing of the hysteretic torque curves.

We would like to also point out a potential difference between our observations in Nb-doped Bi$_2$Se$_3$ and the similar early experimental observation \cite{Zeng} and theoretical explanation \cite{FuPRB2014} of nematic order in the spin susceptibility of Cu-doped Bi$_2$Se$_3$. In the Cu-doped material, quantum oscillations \cite{LawsonPRB,LawsonPRL} and photoemission~\cite{Wray2010,Lahoud} reveal that there is only one bulk Fermi pocket. In contrast, the electronic state in Nb-doped Bi$_2$Se$_3$ shows at least two distinct Fermi surfaces~\cite{LawsonNBS}. Our work calls for further theoretical and experiential exploration of the impact of multi-orbitals on the search for unconventional superconductors.

The unique sensitivity of our in-plane torque magnetometry leads to a promising new tool for probing the superconducting pairing symmetry of other unconventional superconductors. It can elucidate or confirm potential p-wave superconducting symmetry in materials such as Sr$_2$RuO$_4$~\cite{Mackenzie} and UPt$_3$~\cite{StrandUPt3, SaulsUPt3}.  In contrast to those materials, the exponential decay in the heat capacity in Nb-doped Bi$_2$Se$_3$ suggests a superconducting gap without line nodes. It would be very interesting to investigate whether a nodal superconducting gap would enhance or diminish the nematic order in the superconducting state.

\newpage
\newpage
\section*{Materials and Methods}
We preformed torque magnetometry measurements with our home built cantilever setup by glueing the Nb-doped Bi$_{2}$Se$_{3}$ single crystal to the end of a thin beryllium copper cantilever. We then placed the cantilever in an external magnetic field $H$. We measured the torque on the cantilever by tracking the capacitance between the metallic cantilever and a fixed gold film underneath using an AH2700A capacitance bridge with a 7 kHz drive frequency. We calibrated the spring constant of the cantilever by tracking the angular dependence of capacitance caused by the sample weight at zero magnetic field. 
The tilt angle $\phi$ is defined as the angle between the direction of the magnetic field and the positive $x$-axis, which is marked in Fig. 1(A) as the direction of the cantilever arm.

The sample heat capacity is measured in a Quantum Design Physical Properties Measurement Systems (PPMS) using the relaxation method.

The National High Magnetic Field Laboratory provided the magnet and He3 fridge. During the torque magnetometry measurement, samples were soaked in pumped liquid helium 3, and the magnetic field was swept at 0.25 T/min. 

We preformed the magnetization measurement with a Quantum Design Magnetic Properties Measurement System at $H$ = 5 Oe.

{\it Acknowledgements} This work is mainly supported by the Department of Energy under Award No. DE-SC0008110 (magnetization measurement), by the National Science Foundation under Award No. DMR-1255607 (sample growth),  the David and Lucile Packard foundation (theory). Supporting measurements were made possible with the support by the Office of Naval Research through the Young Investigator Prize under Award No. N00014-15-1-2382 (thermodynamic characterization), by the National Science Foundation under Award No. 1307744 (electrical transport characterization), and the National Science Foundation Major Research Instrumentation award under No. DMR-1428226 (supports the equipment of the thermodynamic and electrical transport characterizations).  Some experiments were performed at the National High Magnetic Field Laboratory, which is supported by NSF Cooperative Agreement No. DMR-084173, by the State of Florida, and by the DOE. We are grateful for the assistance of Tim Murphy, Glover Jones, and Ju-Hyun Park of NHMFL. T.A. thanks the Nakajima Foundation for support. B.J.L. acknowledges support by the National Science Foundation Graduate Research Fellowship under Grant No. F031543.

%

\clearpage

\section*{Supplemental Materials}

\subsection{Superconducting Hysteresis}
An external magnetic field kills the superconducting state in a type I superconductor at the critical field, $H_c$. However, in a type II superconductor, there is a mixed state between the lower critical field, $H_{c1}$, and the upper critical field, $H_{c2}$.  For $H_{c1} < H < H_{c2}$, magnetic flux penetrates the superconductor creating a lattice of vortices. Due to defects in the sample, these vortices are pinned in place. In order for the vortices to move, the Lorentz force from a current near the vortices would need to overcome this pinning force. Thus, the pinned magnetic flux has an irreversible response to a changing external magnetic field. This gives rise to hysteresis in the magnetic response of the superconductor.

The Bean model~\cite{Bean62, Bean64} successfully explained this hysteretic feature in type II superconductors. In this model, we assume that the current density in the superconductor can only take the values of 0 and $J_c$, where $J_c$ is the critical current density. Due to the Ampere's law, the spacial profile of $J_c$ determines $\bf{b}(x)$, the magnetic flux density of unit volume at each location $\bf{x}$ in the superconductor. Integrating $\bf{b}(\bf{x})$ gives the total magnetic field density $B$ inside the superconductor. 

\begin{figure}[h]
\includegraphics[width=3.3 in]{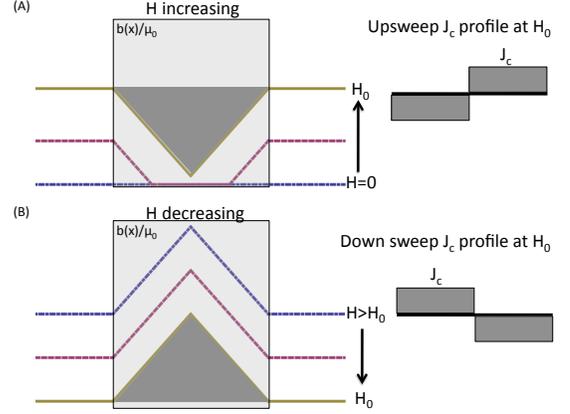}
\caption{\label{fighys} 
{\bf Schematic of magnetic flux density in type II superconductors} (color online)
(A) Magnetic flux density in a type II superconductor as external magnetic field is swept up from $H=0$ to $H_0>0$ according to the Bean model. The right inset shows a sketch of the critical current density profile at $H_0$ during the upsweep. Magnetization $M$ corresponds to the dark shaded area.  (B) Magnetic flux density in a type II superconductor as external magnetic field is swept down from $H>H_{c2}$ to $H_0$. The lagging of the internal magnetic flux density due to flux pinning gives rise to hysteresis in magnetization. The sample's critical current density profile for the down sweep is shown in the right panel. Figure adapted from Ref. ~\cite{LiThesis}.
} 
\end{figure}

Figure \ref{fighys} shows the magnetic flux density inside the superconductor (represented by the grey shaded region) for (A) field increasing from $H = 0$ to $H_0$ and (B) field decreasing to $H_0$ after an applied external field greater than $H_{c2}$. Due to the flux-pinning, the internal magnetic flux density $b(x)/\mu_0$ cannot respond to the change in the sweep direction of the external magnetic field. Fig. \ref{fighys}(A) represents the magnetic flux density $b(x)/\mu_0$ as the applied magnetic field is swept up from zero to $H_0$. In this one dimensional analysis, $J_c  =  \frac{1}{\mu_0}\nabla \bf{b(r)} = \frac{1}{\mu_0}\frac{db(x)}{dx}$ would be simply the slope of $\frac{b(x)}{\mu_0}$ vs. $x$.  The spacial profile of $J_c$ is constant, as shown in the right panel of Fig \ref{fighys}(A).  

The situation of field-sweeping-down is different. As the applied field is swept from a large field to the same $H_0$, the $b(x)/\mu_0$ profile response lags, as show in Fig. \ref{fighys}(B).   

The magnetization of the sample is given by the difference between the average magnetic flux density within the sample, $B$, and the applied field, $H$, outside:
\begin{equation}
M=\frac{B}{\mu_0}-H=\frac{1}{w}\int dx\frac{b(x)}{\mu_0} - H
\label{M2SC}
\end{equation}

where $w$ is the typical width of the domain size in the superconductor, or the sample size if the whole sample is in a single domain. In the case of Panel A, Eq. \ref{M2SC} would be simply the shaded area $-wJ_c$ which gives $M_+$, the magnetization at field sweeping up. 

Furthermore, the difference between Panel A and B demonstrates the hysteresis in the magnetization of type II superconductors. Going through the same analysis as above, we find the magnetization of field sweeping down $M_- = wJ_c$  As a result, the hysteresis loop size $\Delta M (H) \equiv M_+-M_- = 2wJ_c$. Therefore, the measurement of the hysteresis loop is a direct probe of the critical current density in a type II superconductor.

\subsection{Evolution of the nematic order under magnetic fields}

\begin{figure}[h]
\includegraphics[width=3.3 in]{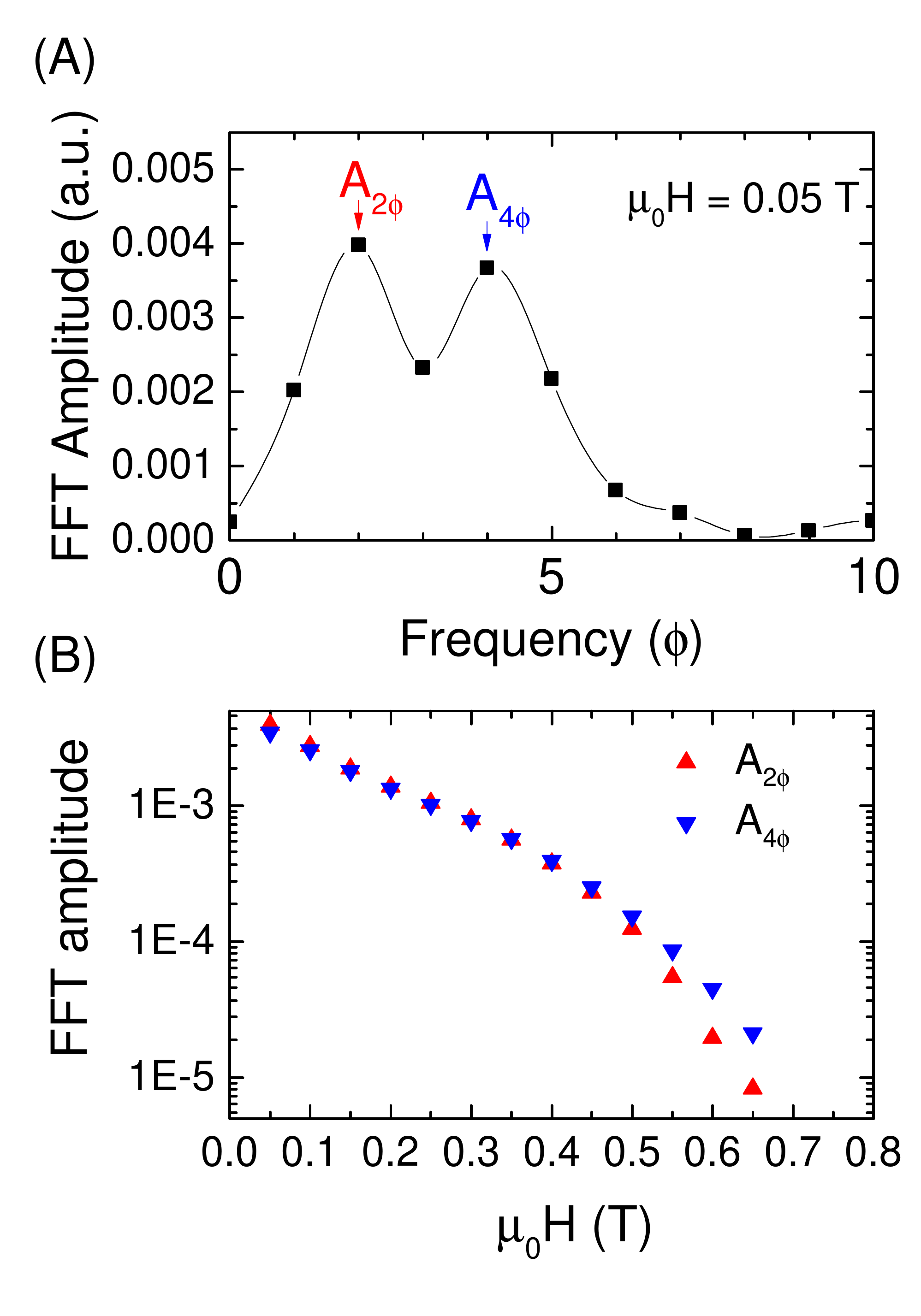}
\caption{\label{figFourier}
{\bf Fast Fourier Transformation of $\Delta M$} (color online)
{(A) Fast Fourier Transform (FFT) of $\Delta M - \phi$ at $\mu_{0}H$ = 0.05 T. The first peak, $A_{2\phi}$, is the amplitude of the nematic order term. The second peak, $A_{4\phi}$, represents the 4-fold term. (B) The magnetic field dependence of the the FFT amplitudes $A_{2\phi}$ and $A_{4\phi}$ in the superconducting state of Nb-doped Bi$_2$Se$_3$. The FFT amplitude is plotted in logarithmic scale for clarify. Above the 0.6 T, the superconducting hysteresis loop quickly vanishes as $H$ approaches the upper critical field. }
} 
\end{figure}

$\Delta M = {M_+-M_-}$ vs $\phi$ for Nb-doped Bi$_2$Se$_3$ is well fitted by the function $f(\phi)=A_{2\phi}sin(2\phi+30^\circ)+A_{4\phi}sin(4\phi-30^\circ)$. We performed Fourier analysis on $\Delta M - \phi$ to determine the strength of the nematic order as a function of magnetic field. Figure \ref{figFourier}(A) shows the Fast Fourier Transform (FFT) of $\Delta M$ at 0.05 T. The first peak, $A_{2\phi}$, is the amplitude of the nematic order $\sin(2\phi)$. The second peak, $A_{4\phi}$, represents the 4-fold term. Because increasing the magnetic field $H$ greatly shrinks the magnetization loop $\Delta M$, both $A_{2\phi}$ and $A_{4\phi}$ dramatically deceases. However, they decrease in different rates. Fig.\ref{figFourier}(B) shows the ratio of $A_{2\phi}$ over $A_{4\phi}$. Both $A_{2\phi}$ and $A_{4\phi}$ are leading terms near zero field. When $H$ increases close to the upper critical field, $A_{2\phi}$ dramatically vanishes. The vanishing of the nematic order as the superconductor approaches the normal state confirms that it is an intrinsic feature of the superconducting state in Nb-doped Bi$_2$Se$_3$.

We would like to note that for a broad field range that absolute amplitude of $A_{2\phi}$ and $A_{4\phi}$ are equal. At this condition, the fitting function turns to be $f(\phi)=2A_{2\phi}sin(\phi-30^\circ)\cos{3\phi}$. This fitting reveals a possible origin of the observed rotational symmetry breaking. The effective superconducting magnetic moment follows the {\it product} of $\sin(\phi-30^\circ)$ and $\cos{3\phi}$, rather than the {\it sum} of two ordering functions. Therefore, there is a strong coupling between the 3-fold crystalline symmetry and a nematic ordering in the triangular superconductor Nb-doped Bi$_2$Se$_3$. The phases between these two sinusoidal functions are locked in the measurement. This observation shows that these two symmetric orderings couple strongly with each other. This suggests that while nematic order is a spontaneous symmetry breaking in the superconducting state, the order seems to be stabilized by the 3-fold crystalline symmetry. The coupling makes sure that the nematic ordering direction is locked to one of the mirror planes of the triangular lattice. As pointed in ref. \cite{FuPRB2014}, the nematic order verifies the two-component nature of the superconducting order parameter. Thus an odd-parity superconducting order is very likely to exist in the ground state, which creates promise for topological superconductivity in Nb-doped Bi$_2$Se$_3$.

\subsection{Symmetry Breaking in different cooling down}

\begin{figure}[ht]
\includegraphics[width=3.3 in]{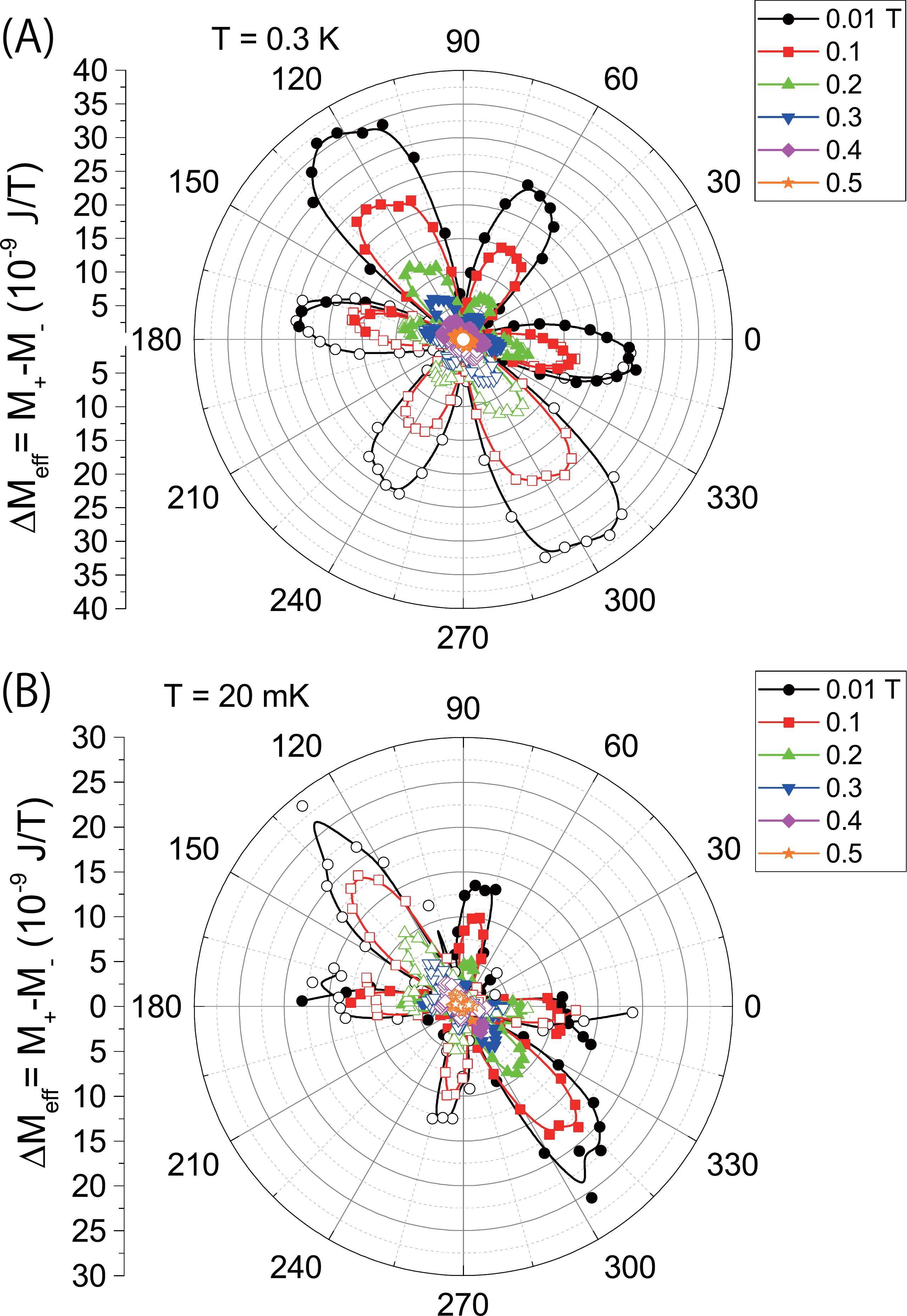}
\caption{\label{figSC}
{\bf $\Delta M = {M_+-M_-}$ vs $\phi$ for two different cool downs.} (color online)
{Polar plot of hysteresis loop magnitude  $\Delta M = {M_+-M_-}$ for two different cool downs. M$_+$ is the magnetization signal from the up-sweep of the magnetic field, and M$_-$ is the magnetization signal from the down-sweep of the magnetic field. (A) Cool down of Nb-doped Bi$_2$Se$_3$ to the 300 mK base temperature in a He-3 cryostat. (B) Cool down of Nb-doped Bi$_2$Se$_3$ to the 20 mK base temperature in a dilution refrigerator.}
}
\end{figure}

Fig.\ref{figSC} shows $\Delta M = {M_+-M_-}$ vs $\phi$ for the same Nb-doped Bi$_2$Se$_3$ in two different cool downs. The data in panel (A) if from a cool down in a He-3 cryostat, and panel (B) is from a different cool down in a dilution refrigerator. In both cases, the preferred axis is locked on a crystalline mirror plane. However, in the two cases, the preferred axis is along roughly the same axis. This second cooling data confirms the observation of the spontaneous symmetry breaking in Nb-doped Bi$_2$Se$_3$.

\subsection{Rotational Symmetry Breaking in other Nb-doped Bi$_2$Se$_3$ samples}

We repeated the search for rotational symmetry breaking in the hysteretic property of Nb-doped Bi$_2$Se$_3$ with a second piece of superconducting Nb-doped Bi$_2$Se$_3$. This sample is designated sample E. Sample E was cooled down to 20 mK in a dilution refrigerator. We swept the field up and down from -1 T to 1 T to measure the entire superconducting hysteresis loop. As in the main text, we applied torque magnetometry to map the complete angular dependence of its in-plane magnetic anisotropy.

\begin{figure}[ht]
\includegraphics[width=3.3 in]{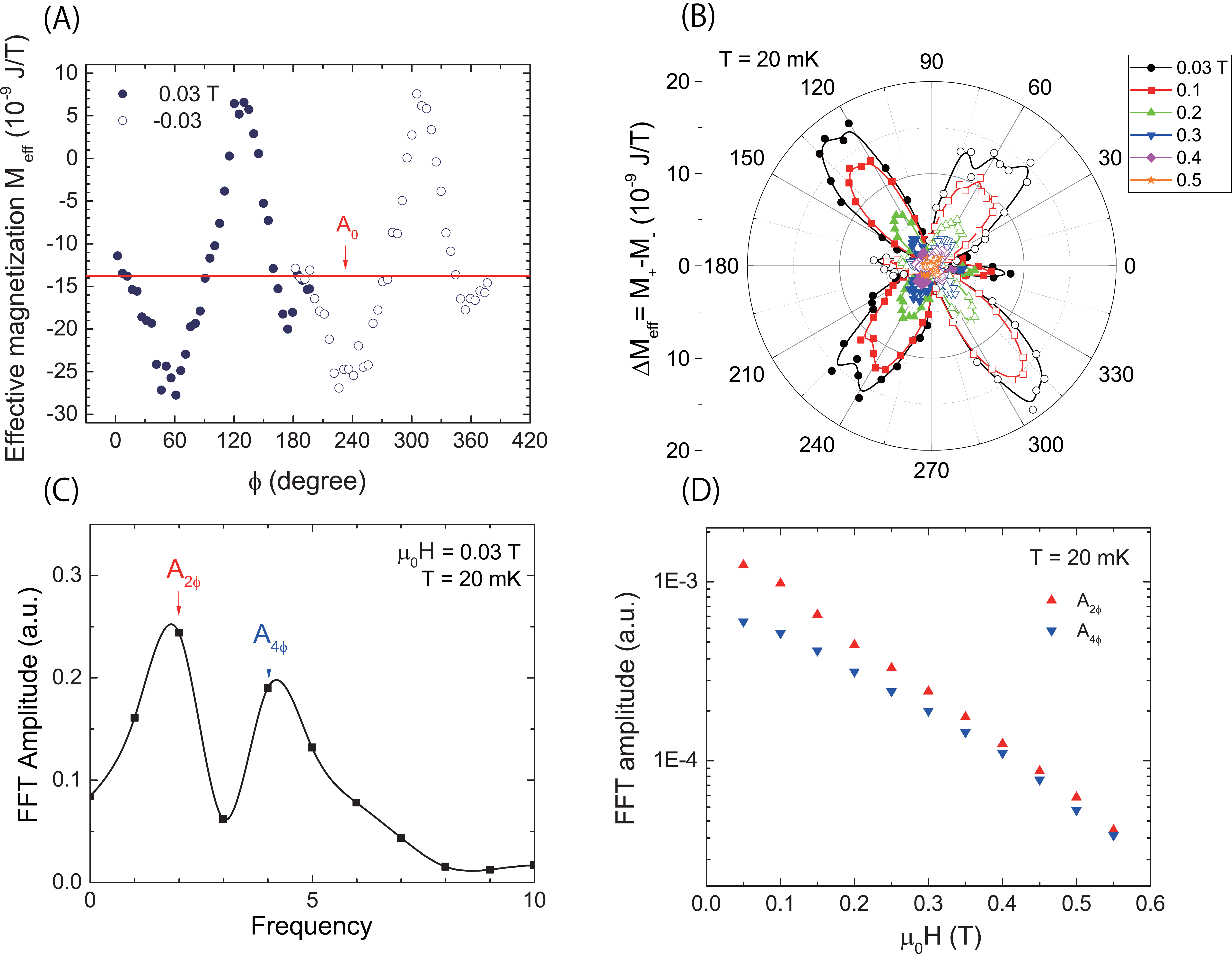}
\caption{\label{figSD}
{\bf Rotational Symmetry Breaking in Sample E} (color online)
{(A) The effective magnetization loop $\Delta M = M_+-M_-$ versus angle $\phi$ for sample E. The red baseline marks a constant offset in the signal that was subtracted out. (B) Polar plot of the effective magnetization in sample E. (C) Here is a Fast Fourier Transform showing the relative strength of the nematic term, $\sin(2\phi)$, and the 4-fold symmetric term, $\sin(4\phi)$. (D) Ratio of the FFT amplitudes $A_{2\phi}$ and $A_{4\phi}$ against external magnetic field in the superconducting state of sample E. Above the 0.6 T, the superconducting hysteresis loop quickly vanishes.}
}
\end{figure}

Fig.\ref{figSD}(A) shows the effective magnetization loop $\Delta M = {M_+-M_-}$ versus angle $\phi$ of sample E. There is a constant background magnetization labeled A$_0$. The effective magnetization loop $\Delta M$ follows $A_0+A_{2\phi}\sin(2\phi-\alpha)+A_{4\phi}sin(4\phi-\beta)$. This angular-independent offset likely arises from the torsional twist of the cantilever setup, although further experiments are needed to determine the exact origin. For further analysis, we subtract away the $A_0$ term to get the angular dependence of the in-plane effective magnetization.
 
 Fig.\ref{figSD}(B) shows a polar plot of the effective magnetization versus angle $\phi$ of sample E. The superconducting hysteresis loop closes at regular intervals corresponding to the axes normal to the mirror planes of the crystal structure. The rotational symmetry is again broken. This is consistent with the sample from the main text. The consistency of the spontaneous symmetry breaking between different superconducting Nb-doped Bi$_2$Se$_3$ samples suggests this is a intrinsic feature of the system and not the result of geometric anisotropy in any given sample.

Fig.\ref{figSD}(C) shows a FFT showing the relative strength of the nematic term, $\sin(2\phi)$, and the crystal symmetry term, $\sin(4\phi)$. The first peak, $A_{2\phi}$, is the amplitude of the nematic order $\sin(2\phi)$. The second peak, $A_{4\phi}$, represents the 4-fold symmetric term.

The ratio $\frac{A_{2\phi}}{A_{4\phi}}$ is a measure of the strength of the nematic order over the background crystal symmetry. As seen in Fig.\ref{figSD}(D), $A_{2\phi}$ is the leading term near zero field. When $H$ increases close to the upper critical field, $A_{2\phi}$ dramatically vanishes. The vanishing of the nematic order as the superconductor approaches the normal state follows the same trend as the main text sample shown in Fig.\ref{figFourier}.

\subsection{Normal State Torque signals Nb-doped Bi$_2$Se$_3$ samples}

\begin{figure}[ht]
\includegraphics[width=3.3 in]{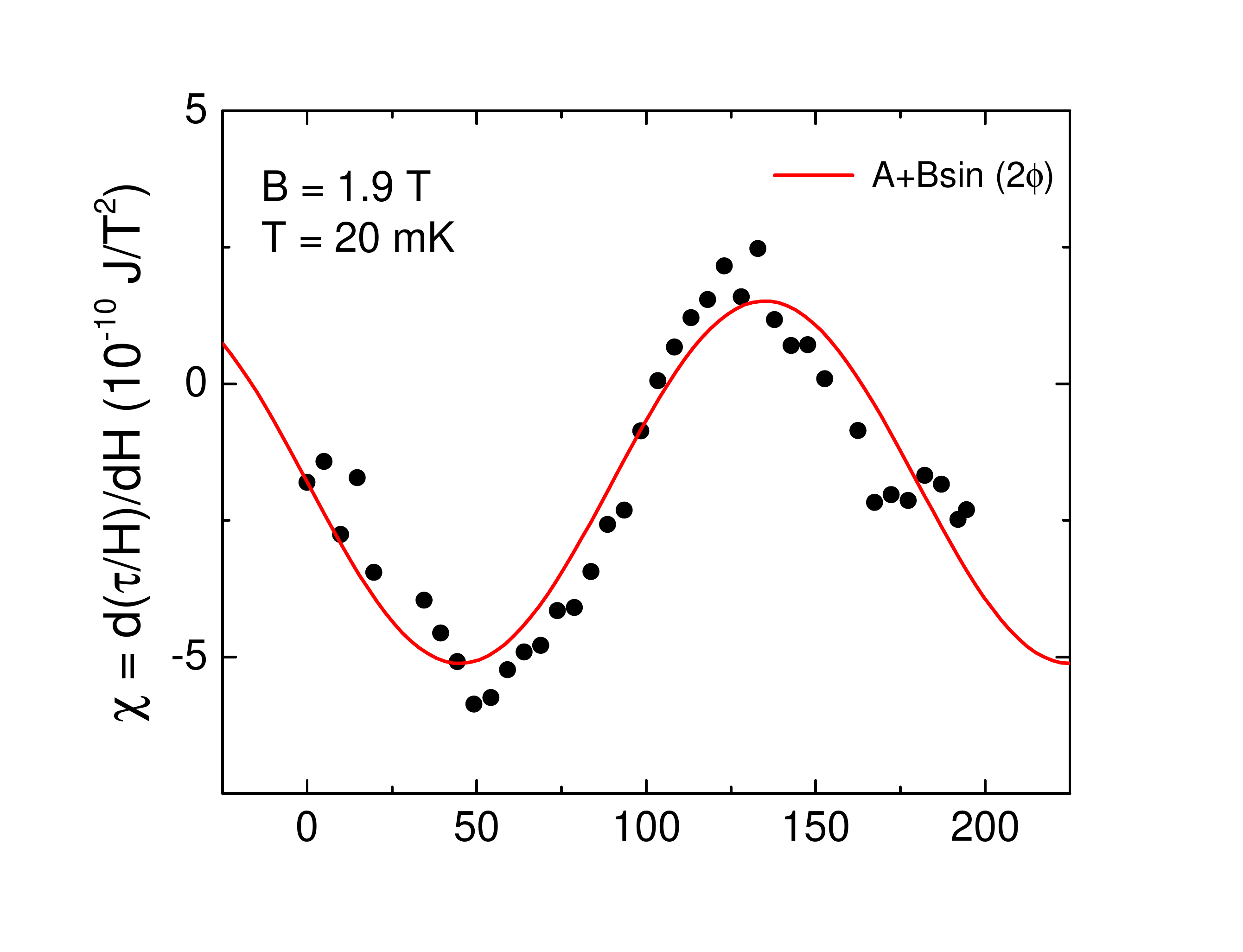}
\caption{\label{figSE}
{\bf Normal state magnetic susceptibility anisotropy of Nb-doped Bi$_2$Se$_3$ in normal state} (color online)
{Magnetic susceptibility anisotropy of Nb-doped Bi$_2$Se$_3$ as a function of angle $\phi$ in external magnetic field above H$_{c2}$. The magnetic susceptibility $dM/dH$ follows a $\sin(2\phi)$ dependence. }
}
\end{figure}

Fig.\ref{figSE} shows the effective magnetic susceptibility $\chi_{eff} = \frac{d(\tau/H)}{dH}$ of Nb-doped Bi$_2$Se$_3$ as a function of angle $\phi$ in $H$ = 1.8 T and 1.9 T external magnetic field. The external field in this figure is much larger than Hc2, thus the sample is not in the superconducting state. Therefore we are able to see in the in-plane anisotropy of the magnetic susceptibility in the normal state. The angular dependence of $\chi_{eff}$  follows the characteristic $\sin{2\phi}$ dependence of the paramagnetic normal state. For a paramagnetic material, the magnetic torque $\tau=\mu_0M \times H$ follows

\begin{eqnarray}
\tau && =\mu_{0}V(\chi_{z}H_{z}H_{x}-\chi_{x}H_{x}H_{z}) \nonumber \\
&& =\mu_{0}V\Delta\chi H^{2}\sin\phi \cos\phi
\end{eqnarray}

where $\Delta\chi=\chi_{z}-\chi_{x}$ is the anisotropy of the magnetic susceptibility of the sample. The observation in Fig. \ref{figSE} shows that the effective susceptibility follows the $\sin{2\phi}$ dependence and the maximum and minimum are aligned with 135 degree and 45 degree, respectively. This observation is consistent with the normal paramagnetic state.


\newpage
\newpage
\newpage



\begin{thebibliography}{99}

\bibitem{RMPHasan}
M. Z. Hasan and C. L. Kane, Rev. Mod. Physics. {\bf 82}, 3045 (2010).

\bibitem{RMPZhang}
S.C. Zhang and X. L. Qi, Rev. Mod. Phys. {\bf 83}, 1057 (2011).

\bibitem{Hor2010}
Y. S. Hor, A. J. Williams, J. G. Checkelsky, P. Roushan, J. Seo, Q. Xu, H. W. Zandbergen, A. Yazdani, N. P. Ong, and R. J. Cava, Phys. Rev. Lett. {\bf 104}, 057001 (2010).

\bibitem{Liu2015}
Zhongheng Liu, Xiong Yao, Jifeng Shao, Ming Zuo, Li Pi, Shun Tan, Changjin Zhang, and Yuheng Zhang, Journal of the American Chemical Society {\bf 137}, 33 (2015).

\bibitem{FuBerg}
L. Fu and E. Berg, Phys. Rev. Lett. {\bf 105}, 097001 (2010). 

\bibitem{Wray2010}
L. A. Wray, S.-Y. Xu, Y. Xia, Y. S. Hor, D. Qian, A. V. Fedorov, H. Lin, A. Bansil, R. J. Cava, and M. Z. Hasan, Nat. Phys. {\bf 6}, 855 (2010).

\bibitem{LawsonPRB}
B.J. Lawson, G. Li, F. Yu, T. Asaba, C. Tinsman, T. Gao, W. Wang, Y.S. Hor, and Lu Li, Phys. Rev. B {\bf 90}, 195141 (2014).

\bibitem{PRLHeatCapacityAndo}
M. Kriener, K. Segawa, Z. Ren, S. Sasaki, and Y. Ando, Phys. Rev. Lett. {\bf 106}, 127004 (2011).

\bibitem{LawsonPRL}
B. J. Lawson, Y. S. Hor, and Lu Li, Phys. Rev. Lett. {\bf 109}, 226406 (2012).

\bibitem{Sasaki}
S. Sasaki, M. Kriener, K. Segawa, K. Yada, Y. Tanaka, M. Sato, and Y. Ando, Phys. Rev. Lett. {\bf 107}, 217001 (2011).

\bibitem{Lahoud} E. Lahoud, E. Maniv, M. S. Petrushevsky, M. Naamneh, A. Ribak, S. Wiedmann, L. Petaccia, Z. Salman, K. B. Chashka, Y. Dagan, and A. Kanigel, Phys. Rev. B {\bf 88}, 195107 (2013).

\bibitem{QiuHor}
Yunsheng Qiu, Kyle Sanders, Jixia Dai, Julia Medvedeva, Weida Wu, Pouyan Ghaemi, Thomas Vojta, and Y.S. Hor, arXiv:1512.03519 (2015).

\bibitem{FuPRB2014}
L. Fu, Phys. Rev. B {\bf 90}, 100509(R) (2014).

%

\bibitem{Zeng}
K. Matano, M. Kriener, K. Segawa, Y. Ando, Guo-qing Zheng, arXiv:1512.07086 (2015).

\bibitem{MatsudaNature}
S. Kasahara et al., Nature (London) {\bf 486}, 382 (2012).

\bibitem{MatsudaScience}
R. Okazaki et al., Science {\bf 331}, 439 (2011).



\bibitem{Sigrist}
M. Sigrist and K. Ueda, Rev. Mod. Phys. {\bf 63}, 239 (1991).

\bibitem{MaoSRO}
Z. Q. Mao, Y. Maeno, S. NishiZaki, T. Akima, and T. Ishiguro, Phys. Rev. Lett. {\bf 84}, 991 (2000).

\bibitem{Hasselbach}
K. Hasselbach, J. R. Kirtley, J. Flouquet, Phys. Rev. B {\bf 47}, 509 (1993).

\bibitem{Movshovich}
R. Movshovich, M. Jaime, J. D. Thompson, C. Petrovic, Z. Fisk, P. G. Pagliuso, and J. L. Sarrao, Phys. Rev. Lett. {\bf 86}, 5152 (2001).

\bibitem{FuTSC2015}
J. W. F. Venderbos, V. Kozii, and Liang Fu, arXiv:1512/04554 (2015).

\bibitem{IshidaYBCO}
T. Ishida, K. Okuda, H. Asaoka, Y. Kazumata, K. Noda, and H. Takei, Phys. Rev. B {\bf 56}, 11897 (1997).

\bibitem{WilleminTBCO}
M. Willemin, C. Rossel, J. Hofer, H. Keller, Z. F. Ren, and J. H. Wang, Phys. Rev. B {\bf 57}, 6137 (1998).


\bibitem{LawsonNBS}
B.J. Lawson, Paul Corbae, Gang Li, Fan Yu, Tomoya Asaba, Colin Tinsman, Y. Qiu, Y.S. Hor, and Lu Li, {\it under review}.

\bibitem{Mackenzie}
A. P. Mackenzie and Y. Maeno, Rev. Mod. Phys. {\bf 75}, 657 (2003).

\bibitem{SaulsUPt3}
J. A. Sauls, Adv. Phys. {\bf 43}, 113 (1994).

\bibitem{StrandUPt3}
J.D. Strand, D.J. Bahr, D.J. Van Harlingen, J.âP. Davis, W.J. Gannon, and W.âP. Halperin, Science {\bf 328}, 1368 (2010).

\end{thebibliography}

\begin{thebibliography}{99}

\bibitem{Bean62}
C. P. Bean, Phys. Rev. Lett. {\bf 8}, 250 (1962).

\bibitem{Bean64}
C. P. Bean, Rev. Mod. Phys  {\bf 36}, 31 (1964).

\bibitem{LiThesis}
Lu Li, "Torque Magnetometry in Unconventional Superconductors", Ph.D. Thesis, Princeton University (2008).


\end{thebibliography}
\end{document}